\documentclass[aps,pra,a4paper,preprint,nofootinbib]{revtex4-1}

\makeatletter
\newcommand\footnoteref[1]{\protected@xdef\@thefnmark{\ref{#1}}\@footnotemark}
\makeatother

\usepackage{amsfonts,amsmath,amssymb,bm,tensor}
\usepackage{comment}
\usepackage{ifpdf}
\usepackage{slashed}
\usepackage{braket}
\usepackage[mathscr]{eucal}
\usepackage[utf8]{inputenc}
\usepackage{cancel}

\ifpdf
\usepackage{graphicx}     
\usepackage[bookmarksopen,colorlinks=true,linkcolor=bblue,citecolor=bblue,urlcolor=ppink]{hyperref}
\else     
\usepackage[dvipdfmx]{graphicx}     
\usepackage[dvipdfmx,bookmarksopen,colorlinks=true,linkcolor=bblue,citecolor=ppink,urlcolor=darkred]{hyperref}
\fi

\usepackage{color}
\definecolor{darkred}{rgb}{0.6,0,0}
\definecolor{darkgreen}{rgb}{0.992447,0.623778,0.034597}
\definecolor{ppink}{rgb}{1,0.4,0.4}
\definecolor{bblue}{rgb}{0.284602,0.317763,0.963947}
\definecolor{mygreen}{rgb}{0,0.7,0}
\definecolor{myred}{rgb}{1,0.3,0.4}
\definecolor{myblue}{rgb}{0.2,0.3,1}

	\newcommand{\km}{\mathrm{km}}

	\newcommand{\g}{\mathrm{g}}
	
	\newcommand{\s}{\mathrm{s}}

	\newcommand{\GeV}{\mathrm{GeV}}
	\newcommand{\TeV}{\mathrm{TeV}}

\newcommand{\bs}{\boldsymbol}
\newcommand{\tx}{\text}

\newcommand{\qcq}{\quad,\quad}
\newcommand{\qcn}{\quad,\nonumber\\}
\newcommand{\df}{\text{d}}
\newcommand{\p}{\partial}
\newcommand{\Mpl}{{M_{pl}}}

\usepackage{enumerate}

\begin{document}


\title{Q-ball decay through A-term in the gauge-mediated SUSY breaking scenario}

\author{Masahiro Kawasaki}
\affiliation{ICRR, University of Tokyo, Kashiwa, 277-8582, Japan}
\affiliation{Kavli IPMU (WPI), UTIAS, University of Tokyo, Kashiwa, 277-8583, Japan}
\author{Hiromasa Nakatsuka}
\affiliation{ICRR, University of Tokyo, Kashiwa, 277-8582, Japan}

\begin{abstract}
	\noindent
	
	Q-balls are non-topological solitons whose classical stability is ensured by a global $U(1)$ charge.
	In particular, Q-balls are produced in the framework of the Affleck-Dine baryogenesis where $U(1)$ charge is the baryon number.
	Since some type of Q-balls is stable against quantum decay into nucleons, it can work as the dark matter.
	When the dark matter Q-balls are captured into the neutron star, they absorb the surrounding neutrons and grow to consume all the neutrons, which leads to a stringent constraint on the Q-ball dark matter.
	However, the Q-ball growth stops due to the $U(1)$ breaking A-term.	
	We revisit the constraint by the neutron star for the gauge-mediation type and the new type Q-ball, and find the allowed parameter space, in which Q-balls work as the dark matter.
	
\end{abstract}

\preprint{IPMU19-0181}
\date{\today}
\maketitle
\tableofcontents

	\section{Introduction}
	\label{sec_thory}


The baryogenesis is a long-standing problem on modern cosmology.
The Affleck-Dine mechanism is one of promising ideas for baryogenesis, which utilizes flat directions in the minimal supersymmetric standard model (MSSM) and can work even in low reheating temperature \cite{Affleck:1984fy,Dine:1995kz}. 
It is known that the flat directions in the MSSM~\cite{Gherghetta:1995dv}, called the Affleck-Dine fields, can also produce non-topological solitons, Q-balls \cite{Coleman:1985ki,Kusenko:1997zq,Kusenko:1997si,Enqvist:1997si,Kasuya:1999wu}.
The properties of Q-balls depend on the potential of the Affleck-Dine field and Q-balls are classified into gauge-mediation type \cite{Kusenko:1997si,Kasuya:2001hg}, gravity-mediation type \cite{Enqvist:1997si,Enqvist:1998en,Kasuya:2000wx}, new type and so on~\cite{Kasuya:2000sc}.
For some parameter space, the Q-balls are stable against decay into nucleons and can explain the dark matter of the Universe~\cite{Kusenko:1997si}. 
Q-balls also have interesting phenomenological applications \cite{Kusenko:2008zm,Kusenko:2009cv,Croon:2019rqu}.  

Previous studies~\cite{Kusenko:1997it,Kusenko:1997vp,Kusenko:2005du} suggest that stability of a neutron star could constrain the abundance of Q-balls in our universe.
When Q-balls collide and stop inside a neutron star, they quickly absorb neutrons and eat up the neutron star, which leads to a stringent constraint on the dark matter Q-balls.
In~\cite{Cotner:2016dhw} (see also \cite{Kasuya:2014ofa}),
however, it is found that the growth of Q-balls is suppressed by the $U(1)$ breaking term (A-term), which inevitably exists in the framework of the Affleck-Dine mechanism.
This is because the A-term breaks the $U(1)$ symmetry and destabilizes the large Q-balls.
The authors in~\cite{Cotner:2016dhw} estimated the destabilization effect by considering the perturbative decay into phonons inside the Q-ball.
However, since the Affleck-Dine field has a large mass outside the Q-balls, the light phonons can not directly escape from the Q-ball and the Q-ball decay needs some extra processes.
So the perturbative decay rate may contain large uncertainties.
Another approach to the Q-ball decay due to the A-term is numerical lattice simulations, which estimate non-perturbative and classical decay rate.
Thus, it is complementary to the perturbative approach.

In this paper, we investigate the instability of Q-balls by numerical lattice simulations.
The numerical estimation for the A-term instability is firstly performed by \cite{Kawasaki:2005xc} on gravity-mediation type Q-balls.
We extend the previous calculation and perform the lattice simulation on the gauge-mediation type as well as gravity-mediation one.

We confirm the instability in both gauge- and gravity-mediation types of Q-balls by the lattice calculation, and put constraints on the size of Q-balls.
Using these results, we perform a comprehensive analysis of the growing Q-balls, which change the type and profile during growth in a neutron star.
We show the allowed parameter space considering the constraints by the neutron star stability.

In Sec.~\ref{sec_potential}, we introduce the Q-balls in the gauge-mediated SUSY breaking. 
The Q-ball growth inside the neutron star is summarized in Sec.~\ref{sec_growQ}.
In Sec.~\ref{sec_nonpertAterm}, we describe the set up of our numerical simulation and show the results on the instability of Q-balls.
Section~\ref{sec_discussion} is devoted to discussion on the evolution of the three types of Q-balls.
We conclude this paper in Sec.~\ref{sec_conclusion}.
The appendix part contains a summary of thin wall type Q-balls  (appendix \ref{sec_eachhistory}) and the calculation of the perturbative Q-ball decay (appendix \ref{sec_pertAterm}).

	\section{Property of Q-balls}
	\label{sec_potential}

In the gauge-mediated SUSY breaking, the AD field has the following low energy effective potential for $|\Phi| \gg M_\text{mess}$ ($M_\text{mess}$: messenger mass)~\cite{Kasuya:2015uka},
\begin{align}
	V(\Phi)
	&= 
	V_\tx{gauge}+V_\tx{gravity}
	 +Am_{3/2}
	 \left( \frac{\Phi^n}{nM_*^{n-3}} + h.c.\right)
	 +	V_\tx{NR}
	 \label{eq:potential}
\\
	 & V_\tx{gauge}
	= M_F^4 \left[
	\log\left( \frac{|\Phi|^2}{M_\tx{mess}^2} \right)
	\right]^2
\\
	& V_\tx{gravity}=
	m_{3/2}^2 |\Phi|^2
	 \left( 
	 1+K\log \left( \frac{|\Phi|^2}{M^2}\right)
	 \right)
\\
	& V_\tx{NR}
	=B \frac{|\Phi|^{2n-2}}{M_*^{2n-6}},
\end{align}
where $K$ is the coefficient of the one-loop correction,
 $m_{3/2}$ is the gravitino mass,
 $M_F$ is the scale of gauge-mediated SUSY breaking\footnote{
 $M_F\gtrsim 5\times 10^5 \GeV$ for the minimal direct gauge mediation model.
 }
 and $M$ is the renormalization scale, which is taken as a typical energy scale of a phenomenon to keep the perturbative description.
$M_*$ is the cut-off scale for the non-renormalizable operator, for example, $M_*\sim10^{16}\GeV$ (GUT scale) or $M_*=\Mpl$ (Planck scale).
We assume that $A$ and $B$ are $\mathcal O(1)$ real parameters.
When the potential $V(\Phi)$ is shallower than quadratic, it has a non-topological soliton (Q-ball) solution~\cite{Coleman:1985ki} which is written as $\Phi(t, r) = \frac{1}{\sqrt{2}}\phi(r) \exp(i\omega t)$ with radial part $\phi(r)$ depending on the potential.

At small field values, the potential is dominated by $V_\tx{gauge}$ which leads to Q-balls called gauge-mediated type. 
The properties of the gauge-mediation type Q-balls are given by~\cite{Kawasaki:2012gk}
\begin{align}
	&R_Q\simeq 
	\frac{1}{\sqrt{2} } 	\zeta^{-1}  	M_F^{-1} Q^{1/4}
,\quad
	\phi_0 
	\simeq
	\zeta
	M_F  Q^{1/4}
,\quad
		\omega 
		\simeq \sqrt{2} \pi \zeta
		M_F Q^{-1/4}
,\quad
	M_Q \simeq 
	\frac{4\sqrt{2}\pi }{3}
	\zeta
	M_F Q^{3/4}
	\label{eq:qball_gauge_med}
\end{align}
where $R_Q$,  $\phi_0$ and $M_Q$ are the size, central field value and mass of the Q-ball, $\zeta$ is $\mathcal O(1)$ parameter \cite{Kawasaki:2012gk}.
The gauge-mediation type Q-balls are stable against decay into nucleons for $\omega < b m_N$, which is achieved when they have the large Q-charge given by~\cite{Kasuya:2015uka}
\begin{align}
	Q>Q_\tx{min}= 
	\left(
	\frac{\sqrt{2} \pi \zeta M_F}{b m_N}
	\right)^4
	\simeq
	1.2\times 10^{30}
	\left(
		\frac{\zeta}{2.5}
		\frac{ M_F}{10^6\GeV}
		\frac{ 1/3}{b }
	\right)^4 
	\ , 
	\label{eq:Qmin}
\end{align}
where $m_N\simeq 1\GeV$ is the mass of a nucleon and $b=1/3$ is the baryon number of the AD field.

For the large field value $\phi \gtrsim \phi_\tx{eq}=\sqrt{2}\frac{M_F^2}{m_{3/2}}$, $V_\tx{gravity}$ dominates the potential and it has Q-ball solutions if $K<0$.
Those Q-balls are generally called gravity-mediation type Q-balls, but in the framework of gauge-mediated SUSY breaking they are called new type.
Taking $M^2=\phi_0^2/2$, the gravity-mediation type Q-balls have the following property: 
\begin{align}
	R_Q=2|K|^{-1/2}m_{3/2}^{-1}\qcq
	\phi_0\sim|K|^{3/4}m_{3/2}Q^{1/2}\qcq
	\omega= m_{3/2}\sqrt{1+2|K|}
	.
	\label{eq:propertygrav}
\end{align}
The stability condition requires $m_{3/2} <  b m_N (1+2|K|)^{-1/2}$.
We also investigate the $V_\tx{gravity}$ with $K>0$. In this case, the $V_\tx{gravity}$ does not support the Q-ball solution and the Q-ball becomes the thin wall type as discussed later.

For an extremely large Q-ball, the potential is lifted by the non-renormalizable term and the field value reaches the critical value $\phi_c$ given by
\begin{align}
	0=\frac{\df }{\df \phi}\left( V/\phi^2\right)\big|_{\phi=\phi_{c}} . 
	\label{eq:criticalcondition}
\end{align}
Such a large Q-ball has the field value $\phi_c$ with a thin surface region and is called thin wall type.
The properties of the thin wall type Q-balls are ~\cite{Coleman:1985ki}
\begin{align}
	R\sim \left(  \frac{3}{4\pi \omega_c \phi_c^2}\right)^{1/3} Q^{1/3}
\qcq
	\phi_0\sim \phi_c
\qcq
	\omega_c^2\sim 2V(\phi_c)/\phi_c^2 .
\end{align}
In the following discussion, the thin wall type Q-ball has following two important properties, (1)the radius grows like $R\propto Q^{1/3}$ as $Q$ increases, and (2) the field value inside the Q-ball no longer increases.
(1) results in a rapid growth rate inside a neutron star and (2) results in the inefficient suppression of its size by the A-term instability, which we discuss later.

Contrary to the $U(1)$ conserving terms, the A-term does not support the Q-ball but destabilizes it.
For the gravity-mediation type, the instability occurs when the A-term becomes comparable to the $V_\tx{gravity}$~ \cite{Kawasaki:2005xc}.
In Sec.~\ref{sec_nonpertAterm}, we investigate the A-term instability based on the ratios of the A-term to the term supporting the Q-ball defined as
\begin{align}
     \xi_\tx{gauge}
	&= \frac{2 |A| m_{3/2} \Phi_0^{n}}{n M^4_F M_*^{n-3}} ,
	\\
	\xi_\tx{gravity}
	& =\frac{2 |A| m_{3/2} \Phi_0^{n-2}}{n m^2_{3/2} M_*^{n-3}}
	.
\end{align}
for gauge- and gravity-mediation type, respectively.

While Q-balls grow inside a neutron star, Q-balls change their type depending on the dominant potential term.
In this paper, we focus on the following three cases:
\begin{itemize}
    \item Case without $V_\tx{gravity}$ 
    ~~~~~~~~~$V_\tx{gauge}\to V_\tx{NR}$,
     \item Case with $V_\tx{gravity}$ ($K<0$) 
    ~~$V_\tx{gauge}\to V_\tx{gravity}(K<0)\to V_\tx{NR}$,
    \item Case with $V_\tx{gravity}$ ($K>0$)  
    ~~$V_\tx{gauge}\to V_\tx{gravity}(K>0)$.
\end{itemize}
For example, in the case with $V_\tx{gravity}$ ($K<0$), Q-balls grow from the gauge-mediation type to the gravity-mediation type.
When they grow more, the $V_\tx{NR}$ prevents the growth of the field value, and their radius starts to grow, which results in the thin wall type. 
In the other case, $K>0$, the $V_\tx{gravity}$ works like the $V_\tx{NR}$, in which the thin wall type appears following the gauge-mediation type.
In this paper, we assume that captured Q-balls are gauge mediation type because the new type Q-balls are not efficiently captured by neutron stars (see next section).

	\section{Growing Q-ball}
	\label{sec_growQ}

Let us summarize the constraint on Q-ball DM from the stability of neutron stars following \cite{Kusenko:2005du,Cotner:2016dhw}.
In this paper we assume that a typical neutron star has the total baryon charge $B_\tx{NS}\sim 10^{57}$,
 the lifetime longer than $\tau_\tx{NS}=0.1\tx{Gyr}$,
 and  the neutron number density $n_0\sim 4\times 10^{-3}\GeV^3$ at the center~\cite{deLavallaz:2010wp,Cotner:2016dhw}. 
Assuming the gauge-mediated type, the flux of the Q-ball DM is estimated as $F\simeq 10^2 Q^{-3/4} \left( \frac{1\TeV}{M_F} \right) \tx{cm}^{-2}\s^{-1}\tx{sr}^{-1}$ \cite{Kusenko:1997vp}. 
For a neutron star with radius $R_\tx{NS}\sim 10\km$, the number of incoming Q-balls is written as
\begin{align}
	 N  &=  4\pi
	\left( \frac{R_\tx{NS}}{1\tx{cm}} \right)^2
	\frac{t}{1\s}
	 10^2 Q^{-3/4} \left( \frac{1\TeV}{M_F} \right) \nonumber \\
	 & \simeq 
	 4.0 \times 10^{4} \frac{t}{1\tx{yr}} 
	 \left( \frac{R_\tx{NS}}{10\km} \right)^2
	\left( \frac{Q}{10^{24}}\right)^{-3/4}
	 \left( \frac{10^{3}\GeV}{M_F} \right) .
\end{align} 
For the cosmological time scale ($t=\mathcal O(\tx{Gyr})$), a significant number of Q-balls collide into a neutron star.
The capture of gauge-mediation type Q-balls by a neutron star is discussed in~\cite{Kusenko:1997it}.
They describe the motion of a gauge-mediation type Q-ball inside the neutron star by the equation, $\ddot x = -\Omega^2 x -\gamma \dot x$, where $x$ is the position of the Q-ball from the center of the neutron star,   $\Omega^2=(4\pi\rho_\tx{ns})/ (3\Mpl^2)$ is the angular frequency of the orbital motion and $\gamma\sim \pi R_Q^2\rho_\tx{ns} v_n/M_Q $ is the friction inside the neutron star.
Here  $\rho_\tx{ns}$ is the matter density of the neutron star and $v_n$ is the speed of the neutrons in the star, where we assume $v_n\sim 1$ in this paper.
Q-balls are captured by the neutron star when the friction dominates the angular motion as
\begin{align}
	1\ll 
	\frac{\gamma}{\Omega}
	&=
		 \frac{3\rho_\tx{ns} v_n }{\zeta^3 8 \sqrt{2} } M_F^{-3}  Q^{-1/4}
		\sqrt{\frac{3}{4\pi}\frac{M_{pl}^{\,2}}{\rho_\tx{ns} } }
	\nonumber
\\&	=
		1.3
		\left(\frac{n_\tx{ns}}{4\times10^{-3}\GeV^3 }\right)^{1/2}
		v_n
		\left(\frac{2.5}{\zeta} \right)^3
		\left( \frac{10^3\GeV}{M_F  } \right)^3
		\left( \frac{10^{24}  }{ Q  } \right)^{1/4}
		\label{eq:Qcapture}
\end{align} 
The light Q-balls with small $Q$ easily stop inside the neutron star and Q-balls with large $M_F$ rarely stop.
For the minimal stable Q-ball with $Q_\tx{min}$ given by Eq.\eqref{eq:Qmin}, the efficiency of the capture rate Eq.\eqref{eq:Qcapture} is converted to the condition on $M_F$ as
\begin{align}
	M_F<M_{F,\tx{capture}}
	&=2.5\times 10^3 \GeV
	~
	\frac{2.5}{\zeta}
	v_n^{1/4}
	\left( \frac{4\times 10^{-3}\GeV^3}{n_\tx{ns}} \right)^{3/8}
\end{align}

We comment on the new type Q-ball as DM.
If the DM Q-ball is the new type with $K<0$, the above estimation of the capture rate is modified since the new type has the constant radius independent of its Q-charge as Eq.\eqref{eq:propertygrav}. 
In this case, $\gamma/\Omega$ is smaller than that of the gauge mediation type and written as
\begin{align}
    \frac{\gamma}{\Omega}
    &=\pi \rho_\tx{ns} v_n
    4|K^{-1}|m^{-3}_{3/2} Q^{-1}
	\sqrt{\frac{3}{4\pi}\frac{M_{pl}^{\,2}}{\rho_\tx{ns} } }
\nonumber\\
    &=
	0.03\times 
	\left(\frac{n_\tx{ns}}{4\times10^{-3}\GeV^3 }\right)^{1/2}
	v_n
	\left( \frac{0.1\GeV}{m_{3/2}}  \right)^3
	\frac{0.03}{|K|}
	\frac{10^{24}}{Q}
	,
	\label{eq:captureeffGauge}
\end{align}
which is smaller than that of the gauge mediation type.
Thus, the capture of the new type Q-balls is inefficient [see Fig.\ref{fig:comparegrowt}] and hence we assume that only gauge-mediation type Q-balls are captured inside neutron stars in this paper.

After a Q-ball stops inside a neutron star, 
it starts to absorb the neutrons around it since the energy per baryon number is lower inside Q-balls than outside~\cite{Kusenko:1997vp,Kusenko:2004yw}.  
The growth rate of the Q-ball in a neutron star has large uncertainty. 
Following~\cite{Cotner:2016dhw}, we briefly summarize two estimations of the Q-ball growth rate; one is based on the surface conversion and the other is hydrodynamic consideration.
In the surface conversion scenario, we simply assume that all neutrons are absorbed into Q-balls when they collide on the surface of Q-balls. 
The growth rate is given by
\begin{align}
	\dot Q_\tx{grow}
	&=
	b^{-1}4\pi R^2 n_0 v_n
	\sim
	0.1 \GeV \left(\frac{R}{\GeV^{-1}}\right)^2.
\end{align}

In the hydrodynamic consideration scenario, authors in \cite{Kusenko:2005du} assume that the excess energy from absorbed neutrons produces the pion cloud around the Q-ball. 
Based on the balance of the pressure between the neutron and the pion cloud, the growth rate is evaluated as \cite{Kusenko:2005du},
 \begin{align}
 	\dot Q_\tx{grow}
 	\sim 
 	\frac{10\pi^{3/2} n_0^{5/6} R^2 }{b\sqrt{3\tau}}
 	\sim 1.0\times 10^{-4}\, \GeV \left(\frac{R}{\GeV^{-1}}\right)^2 , 
 	\label{eq_QgrowRate}
 \end{align}
where $\tau\sim 10^8\GeV^{-1}$ is the neutral pion lifetime.
We adopt Eq.\eqref{eq_QgrowRate} as the conservative growth rate in this paper.

Since the growth rate depends on the surface area $4\pi R^2$, the Q-ball growth is determined by the $R$-$Q$ relation of each type.
Using Eq.\eqref{eq_QgrowRate}, the Q-ball grows as
 \begin{align}
	Q(t) & \sim \left(Q_\tx{in}^{1/2}+ 2\times 10^{14}\left(  \frac{2.5}{\zeta}	  \right)^2
	\left( \frac{10^6\GeV}{M_F}  \right)^2   \frac{t}{\tx{yr}}  \right) ^2 
	~~~~\tx{(gauge-mediation type)},
	\label{eq:growtingGaugeQ}
\\
   Q(t) & \sim Q_\tx{in}
	+ 6\times 10^{31}\left( \frac{0.03}{|K|} \right)
	\left( \frac{0.1\GeV}{m_{3/2}} \right)^2
	 \frac{t}{\tx{yr}}  
	 ~~~~~~~~~~~~~\tx{(gravity-mediation type)}
	 \label{eq:growtingGravQ}  ,
\\
    Q(t) & \sim 
	 \left(Q_\tx{in}^{1/3}
	 + 5\times 10^{26} 
	 \left(   \frac{1 \GeV^6 }{ V(\phi_c) \phi_c^2 } \right)^{1/3}\frac{t}{\tx{yr}}  
	 \right) ^3 
	 ~~~~~~~~~~~~\tx{(thin wall type)}.
	 \label{eq:thinwallGrowthrate}
 \end{align} 
\begin{figure}[t]
	\includegraphics[width=1.\textwidth]{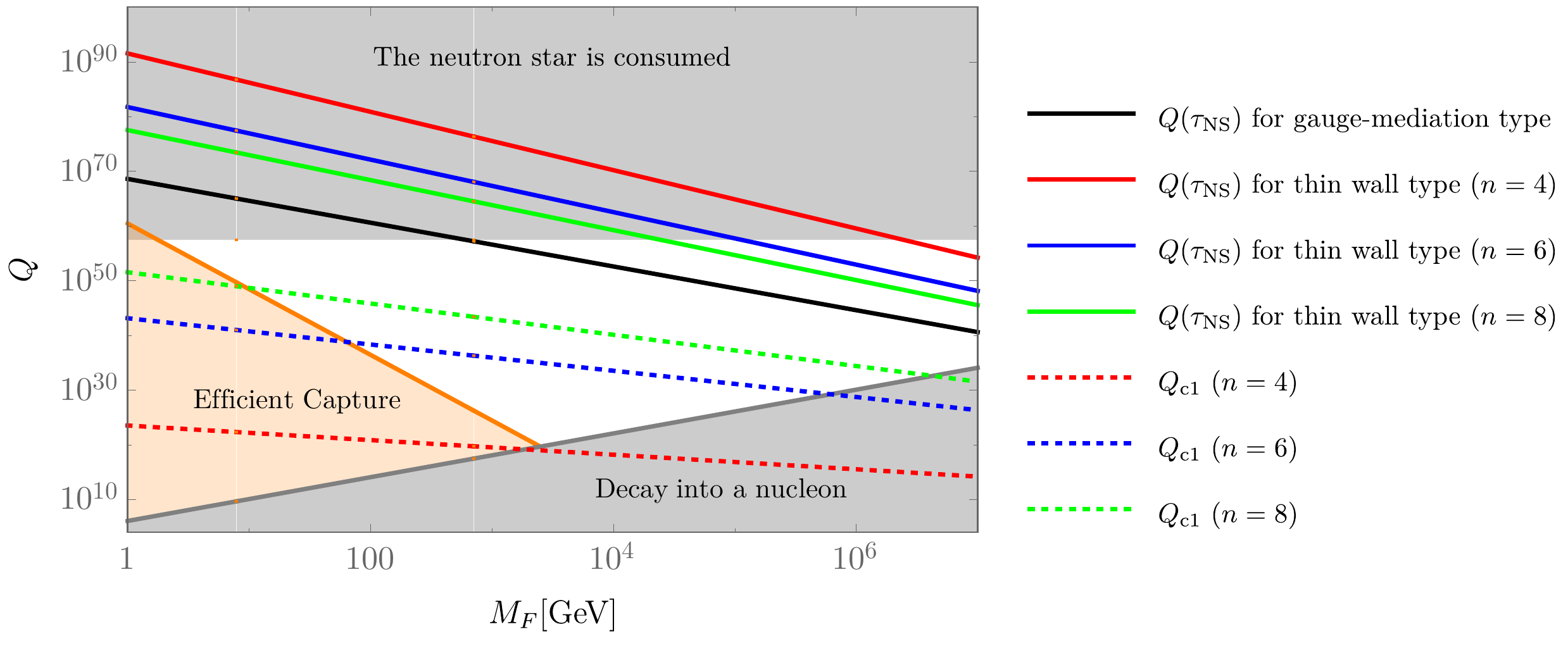}
	\caption{
	    The black solid line is Q-ball charge after it evolves over $\tau_\tx{NS}=0.1\tx{Gyr}$ for the gauge-mediation type [Eq.\eqref{eq:growtingGaugeQ}].
	    Since the growing Q-ball overcomes the threshold value of the thin wall type $Q_{c1}$ (dotted lines) given by Eq.\eqref{eq:gaugeTranceQc1}, $Q(\tau_\tx{NS})$ should be evaluated using the growth rate of thin wall type [see Eq.\eqref{eq:thinwallGrowthrate}] (red, blue, green lines for n=4,6,8).
	    For the small $M_F$, the Q-ball consume all the baryon in the neutron star.
	}
	\label{fig:GrowthofQball}
\end{figure}
Let us consider the typical evolution of a Q-ball inside a neutron star without $V_\tx{gravity}$ and A-term, for simplicity.
For $M_F\lesssim \mathcal O(10^3)\GeV$, a stable DM Q-ball with $Q\gtrsim Q_\tx{min}$ satisfies the condition of efficient capture (orange region in Fig.\ref{fig:GrowthofQball}) given by Eq.\eqref{eq:Qcapture}.
At first, the Q-ball follows the growth rate of the gauge-mediation type Eq.\eqref{eq:growtingGaugeQ}.
During the growth, the Q-ball becomes the thin wall type and more quickly grows than the gauge-mediation type. 
The dotted lines in Fig.\ref{fig:GrowthofQball} describe the threshold value of the thin wall type $Q_{c1}$ given by Eq.\eqref{eq:gaugeTranceQc1}, which is smaller than the  $Q(\tau_\tx{NS})$ for gauge-mediation type (black solid line).
The thin wall type Q-ball quickly grows (color solid lines) and consume all the baryon in the neutron star for $M_F\lesssim \mathcal O(10^4-10^6)\GeV$ depending on the index of the non-renormalizable term.

Note that the time scale of Q-ball growth is much longer than that for the Q-ball to sink inside the neutron star $t= \gamma^{-1}=\mathcal O(10^{-12}\tx{yr}) (Q/10^{24})^{1/4}(M_F/(10^3\GeV))^{3}$. 
Thus, in this paper, we simply assume that only one large Q-ball stays at the center of the neutron star.
    
Now we include the decay of Q-balls through the A-term. 
When the A-term starts to destabilize the Q-ball, the Q-ball size reaches its upper bound $Q_\tx{max}$. 
After the Q-balls stop the growth, the baryon consumption rate in Eq.\eqref{eq_QgrowRate} is determined by its maximum radius $R_\tx{max}$. 
In the next section, we evaluate $Q_\tx{max}$ by the numerical calculations.
 
As a conservative constraint, we require that the Q-balls should not consume all baryon of the neutron star over $\tau_\tx{NS}$.
When the A-term is inefficient, the Q-ball continue to grow until $\tau_\tx{NS}$, which leads to the following constraint: 
\begin{align}
	B_\tx{NS} >
    bQ(\tau_\tx{NS}).
\end{align}
where $Q(\tau_\tx{NS})$ is given by Eqs (\ref{eq:growtingGaugeQ})-(\ref{eq:thinwallGrowthrate})
On the other hand, when the A-term strongly suppresses the Q-ball growth, it quickly reaches the maximum size $Q_\tx{max}$ and the baryon consumption rate $\dot Q_\tx{grow}(R_\tx{max})$ balances to the A-term decay.
Assuming that the growth is quicker than $\tau_\tx{NS}$, the constraint is approximately given by
\begin{align}
	B_\tx{NS} \gtrsim
    b\tau_\tx{NS} \dot Q_\tx{grow}(R_\tx{max})  .
	\label{eq:Bconsumed}
\end{align}
Let us convert $B_\tx{NS} > b \tau_\tx{NS} \dot Q_\tx{grow}(R_\tx{max}) $ into the constraint on $Q_\tx{max}$.
For gauge-mediation type, from Eqs.~(\ref{eq:qball_gauge_med}) and (\ref{eq_QgrowRate}) we obtain
\begin{align}
	Q_\tx{max}^\tx{(gauge)}  & < 
	B_\tx{NS}^2
	\left[  
		b \times10^{-4} \GeV \times 0.1\tx{Gyr}
		\frac{1}{2\zeta^2} 
		\left( \frac{1\GeV}{M_F}\right)^2
	\right]^{-2}
	\nonumber \\
	 &=6\times 10^{69} 	
	\left( \frac{B_\tx{NS}}{10^{57}}\right)^2
	\left( \frac{M_F}{10^6\GeV} \frac{\zeta}{2.5}\right)^4
	\label{eq:QmaxGauge}
\end{align}
For gravity mediation type, $b \tau_\tx{NS} \dot Q_\tx{grow}^\tx{(gravity)}$ is independent of $Q_\tx{max}$. Then, we can convert $B_\tx{NS} > b \tau_\tx{NS} \dot Q_\tx{grow}(R_\tx{max}) $ into the constraint on $m_{3/2}$ as
\begin{align}
	m_{3/2} 
	& >1\GeV \times
	B_\tx{NS}^{-1/2} 
	\left[  
		b \tau_\tx{NS} \ 10^{-4} \GeV  
		\frac{4}{|K|}
	\right]^{1/2} \nonumber \\
	 & = 
	1\times 10^{-10}\GeV 
	\left( \frac{10^{57}}{B_\tx{NS}} \frac{0.03}{|K|} 
	\frac{\tau_\tx{NS}}{0.1\tx{Gyr}}
	\right)^{1/2}
	\label{eq:Gravm32upperconsumption}
\end{align}
In the following section, we investigate the upper bound of Q-charges suppressed by the A-term.

	\section{Numerical simulation of A-term decay}
	\label{sec_nonpertAterm}

We investigate the A-term destabilization of Q-balls using numerical lattice simulation.
Let us describe the setup for calculation.
We initially put a Q-ball in the box and follow the time evolution with the A-term.
The initial profile is based on the gauge or gravity-mediation type with small random fluctuations of $\mathcal O(10^{-6})$.
We expand the complex field by the two real field $f$ and $g$ as $\Phi = \frac{1}{\sqrt{2}}\phi_0 (f+ig)$, where $\phi_0$ is the field value at the center of the Q-ball. 
The lattice simulation is performed for both gauge and gravity-mediation types.
The Q-ball evolution is studied mainly in simulations with 1 spatial dimension. 
We also perform 2 dimensional simulations in some cases and confirm that our results are independent of the spatial dimension.
We use the periodic boundary condition and check that the box size and the resolution do not change the results.

We calculate the integrated energy and Q-charge around the Q-ball as~\cite{Cotner:2016dhw,Coleman:1985ki}
\begin{align}
	E&=\int^{|\bs x|< R_*} \df^3\bs x (\p_\mu \Phi^\dag \p^\mu \Phi +V(\Phi))
	\ , 
\quad
	Q
	=i \int^{|\bs x|< R_*} \df^3 \bs x (\dot\Phi^\dag \Phi-\Phi^\dag \dot\Phi),
	\label{eq:EQdefinition}
\end{align}
where $R_*$ is the radius where the local energy density is smaller than 1\% of that at the center of the Q-ball.

	\subsection{The setup for the gauge-mediation type}
	\label{sec_setupgauge}

We use the following simplified potential:
\begin{align}
	V(\Phi)
	&= 
	M_F^4 \log\left(1+\frac{|\Phi|^2}{M_\tx{mess}^2} \right)
	+Am_{3/2}\left( \frac{\Phi^n}{nM_*^{n-3}} + h.c.\right)
	+\alpha \frac{|\Phi|^{2n-2}}{M_*^{2n-6}}
	.
\end{align}
For convenience in the numerical calculation, we modified $V_\tx{gauge}$ by adding 1 in the argument of the logarithm and using the linear logarithmic potential instead of the quadratic one.
Since the A-term is effective for the large field value, our modification does not change the typical property of gauge-mediation type Q-balls. 
We also add small higher-order term to avoid unstable directions in the potential.

We normalize the space-time coordinates as  $\hat x^\mu=M_\phi x^\mu$ by using the effective mass $M_\phi^2= M_F^4/M_\text{mess}^2$.
The equation of motions for the real and imaginary parts of $\Phi = \frac{1}{\sqrt{2}}\phi_0 (f+ig)$ are given by
\begin{align}
0&= 
	\left( 
	\hat \p_{ t}^2  
	-\hat \nabla^2 
	+\frac{1 }{1+u (f^2+g^2) }
	+
	\tilde \alpha 
	(f^2+g^2)^{n-2} 
	\right)
	\begin{pmatrix}
	f\\g 
	\end{pmatrix}
\\&\quad
	+\frac{n \xi_\tx{gauge}}{4u}
	\begin{pmatrix}
	(f-ig)^{n-1}+(f+ig)^{n-1}
	\\-i (f-ig)^{n-1}  + i (f+ig)^{n-1} 
	\end{pmatrix},
\end{align}
where $u=\frac{1}{2}\phi_0^2/M_{mess}^2$ and $\xi_\tx{gauge}$ is the ratio of the A-term to the logarithmic one given by
\begin{align}
	 \xi_\tx{gauge}
	&= \frac{2 |A| m_{3/2} \Phi_0^{n}}{n M^4_F M_*^{n-3}} .
\end{align}
We take $u=5\times 10^3$ and $45\times 10^3$ to focus on the gauge-mediation type.

	\subsection{The setup for the gravity-mediation type}
	\label{sec_setupgrav}

The potential for the gravity-mediation type is written as
\begin{align}
	V(\Phi)
	&= 
	m_{3/2}^2 |\Phi|^2
	\left( 
    	1+K\log \left( \epsilon +\frac{|\Phi|^2}{M^2}\right)
	\right)
	+Am_{3/2}\left( \frac{\Phi^n}{nM_*^{n-3}} + h.c.\right)
	+\alpha\frac{|\Phi|^{2n-2}}{M_*^{2n-6}}
	\ , 
\end{align}
where we have added a small regulator $\epsilon$ as $|K\log\epsilon|\ll 1$ to avoid the singular behavior at $\phi\sim 0$.
We have chosen the renormalization scale as $M^2=\tfrac{1}{2}\phi_0^2$ to keep the perturbative analysis at the center of the Q-ball.

With $\hat x^\mu= m_{3/2} x^\mu$, the equations of motions for the real and imaginary parts of $\Phi = \frac{1}{\sqrt{2}}\phi_0 (f+ig)$ are written as
\begin{align}
	0&= 
	\left( 
	\hat \p_{ t}^2  
	-\hat \nabla^2 
	+
	1+K\frac{f^2+g^2}{\epsilon + f^2+g^2 }
	+K\log\left( \epsilon+f^2+g^2 \right)
	+\tilde \alpha(f^2+g^2)^{n-2}
	\right)
	\begin{pmatrix}
	f\\g 
	\end{pmatrix}
\nonumber
\\&	\quad
	+\frac{n \xi_\tx{gravity}}{4}
	\begin{pmatrix}
	(f-ig)^{n-1}+(f+ig)^{n-1}
	\\i^{-1} (f-ig)^{n-1}  - i^{-1} (f+ig)^{n-1}
	\end{pmatrix},
\end{align}
where $\xi_\tx{gravity}$ is the ratio of the A-term to the quadratic one as
\begin{align}
	\xi_\tx{gravity}
	=\frac{2 |A| m_{3/2} \Phi_0^{n-2}}{n m^2_{3/2} M_*^{n-3}}
	.
\end{align}

	\subsection{Result}
	\label{sec_numresult}

\begin{figure}[t]
\begin{tabular}{c}
	
	\begin{minipage}{0.5\hsize}	\begin{center}
			\includegraphics[width=1.\textwidth]{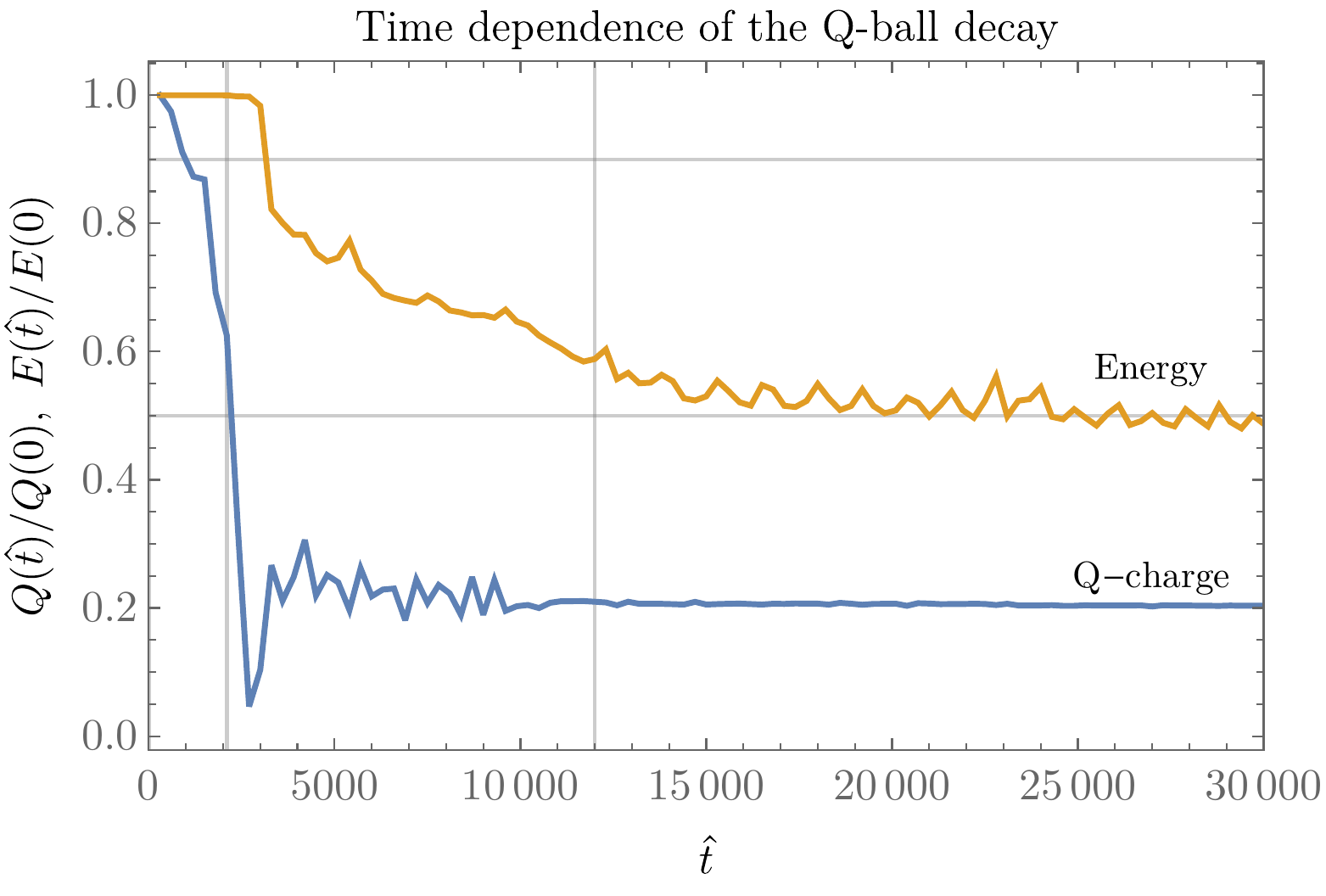}
	\end{center}\end{minipage}
	
	\begin{minipage}{0.5\hsize}	\begin{center}
			\includegraphics[width=1.\textwidth]{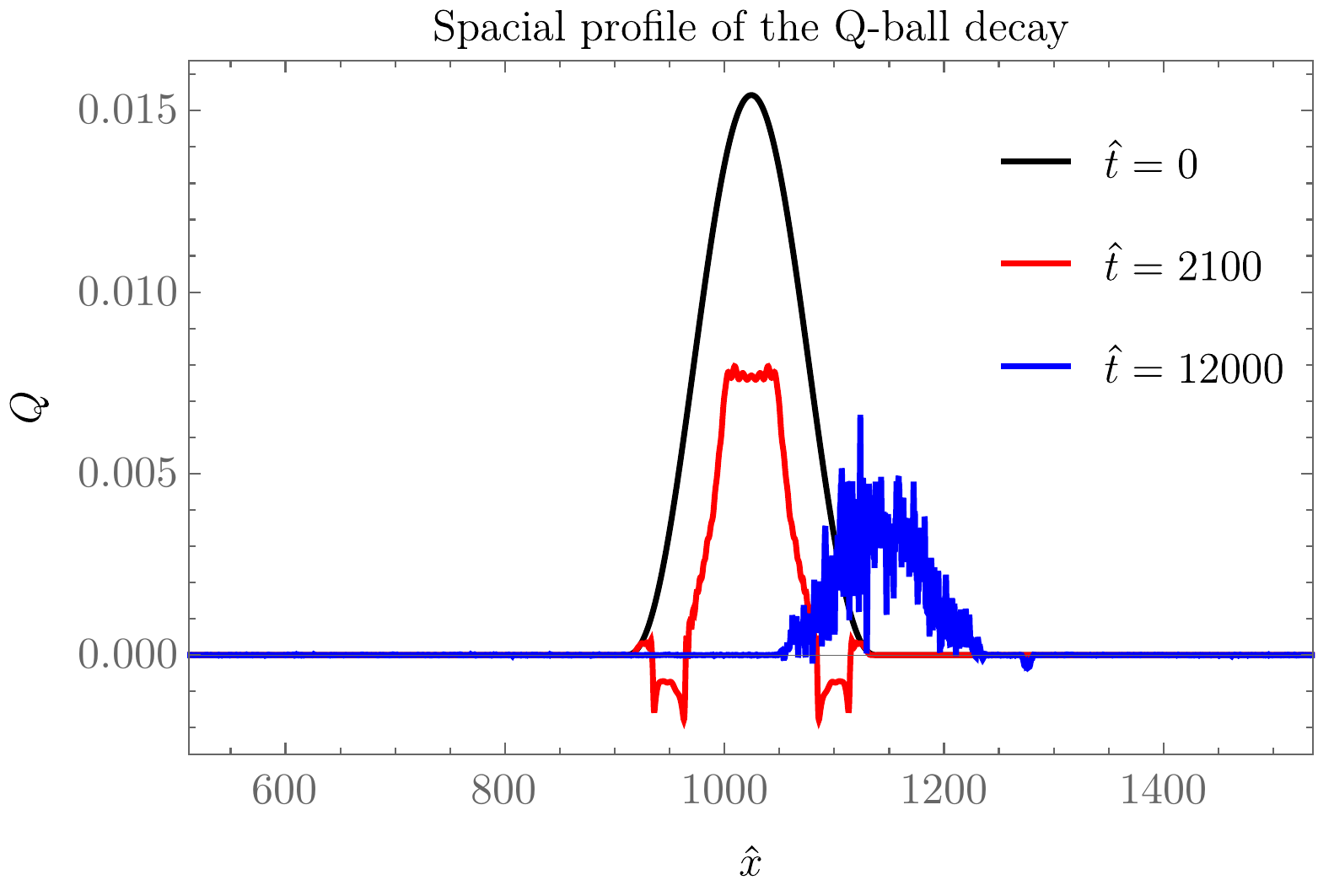}
	\end{center}  \end{minipage}
\end{tabular}
	\caption{
		A typical example of a Q-ball decay in 1 dimensional case.
		The Q-ball is gauge-mediation type with $n=6$, $u=45\times 10^3$ and $\xi_\tx{gauge}=2.25$.
		In the left figure, the energy and Q-charge around Q-ball [Eq.\eqref{eq:EQdefinition}] are normalized by the initial value, $Q(\hat t)/Q(0)$ and $E(\hat t)/E(0)$.
		In the right figure, we plot spatial profiles of a local Q-charge at three $\hat t$.
	}
	\label{fig:ExamplesOfQballDecays}
\end{figure}

\begin{figure}[t]
	\begin{tabular}{c}
		
		\begin{minipage}{0.3\hsize}	\begin{center}
				\includegraphics[width=1.\textwidth]{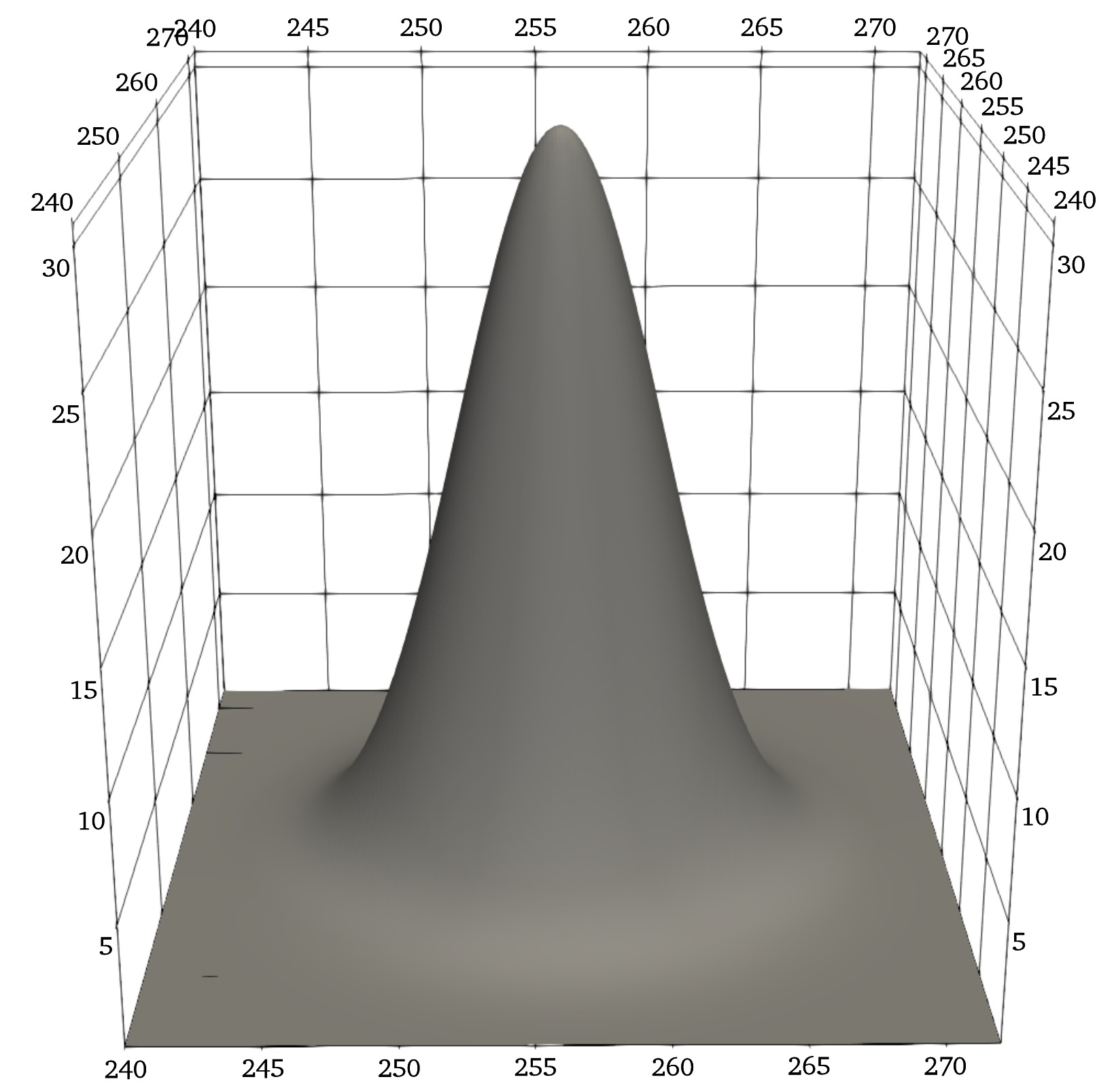}
				$\hat t=0$
		\end{center}\end{minipage}
		
		\begin{minipage}{0.3\hsize}	\begin{center}
		\includegraphics[width=1.\textwidth]{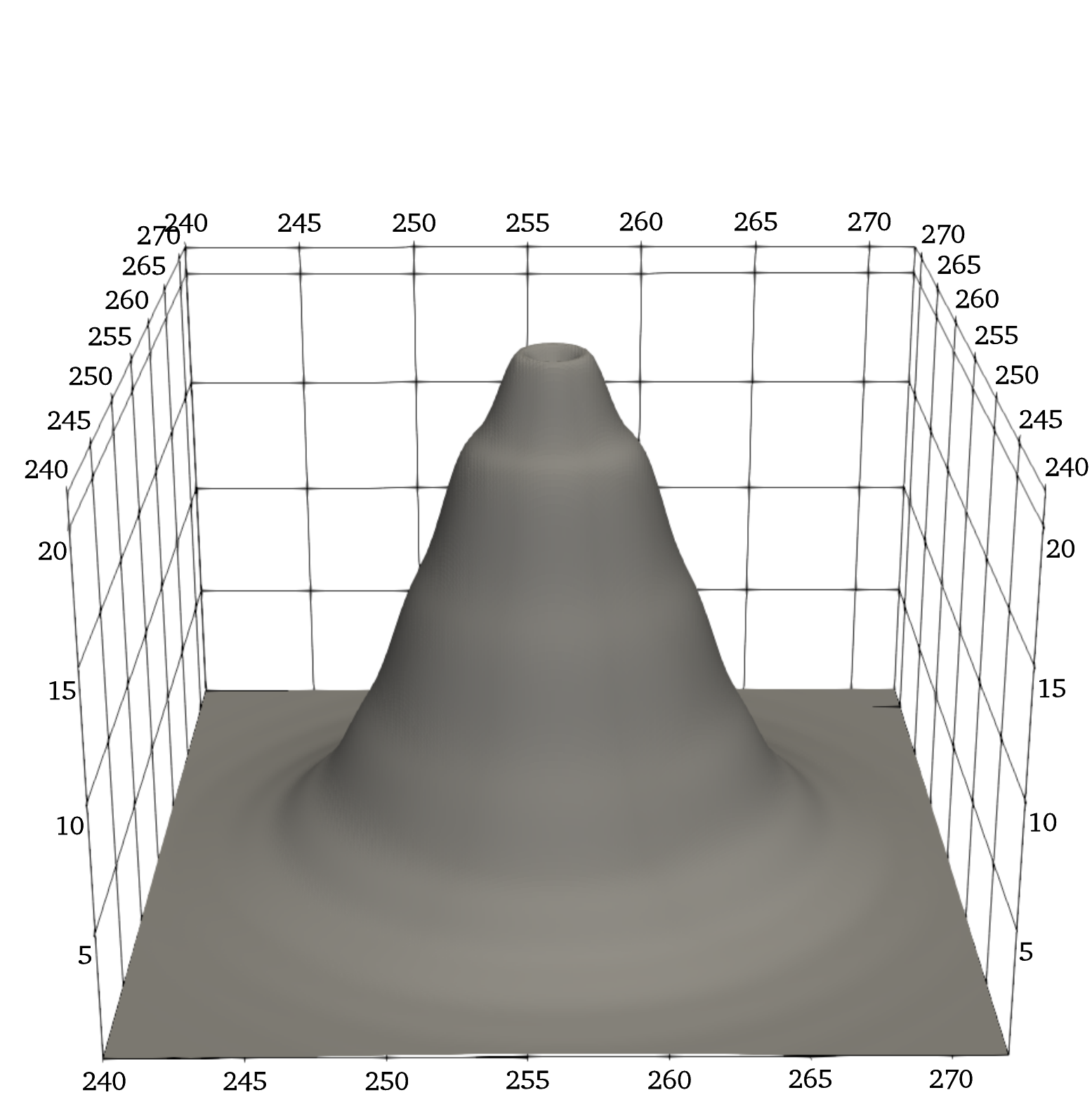}
				$\hat t=6\times10^3$
		\end{center}  \end{minipage}
	
		\begin{minipage}{0.3\hsize}	\begin{center}
		\includegraphics[width=1.\textwidth]{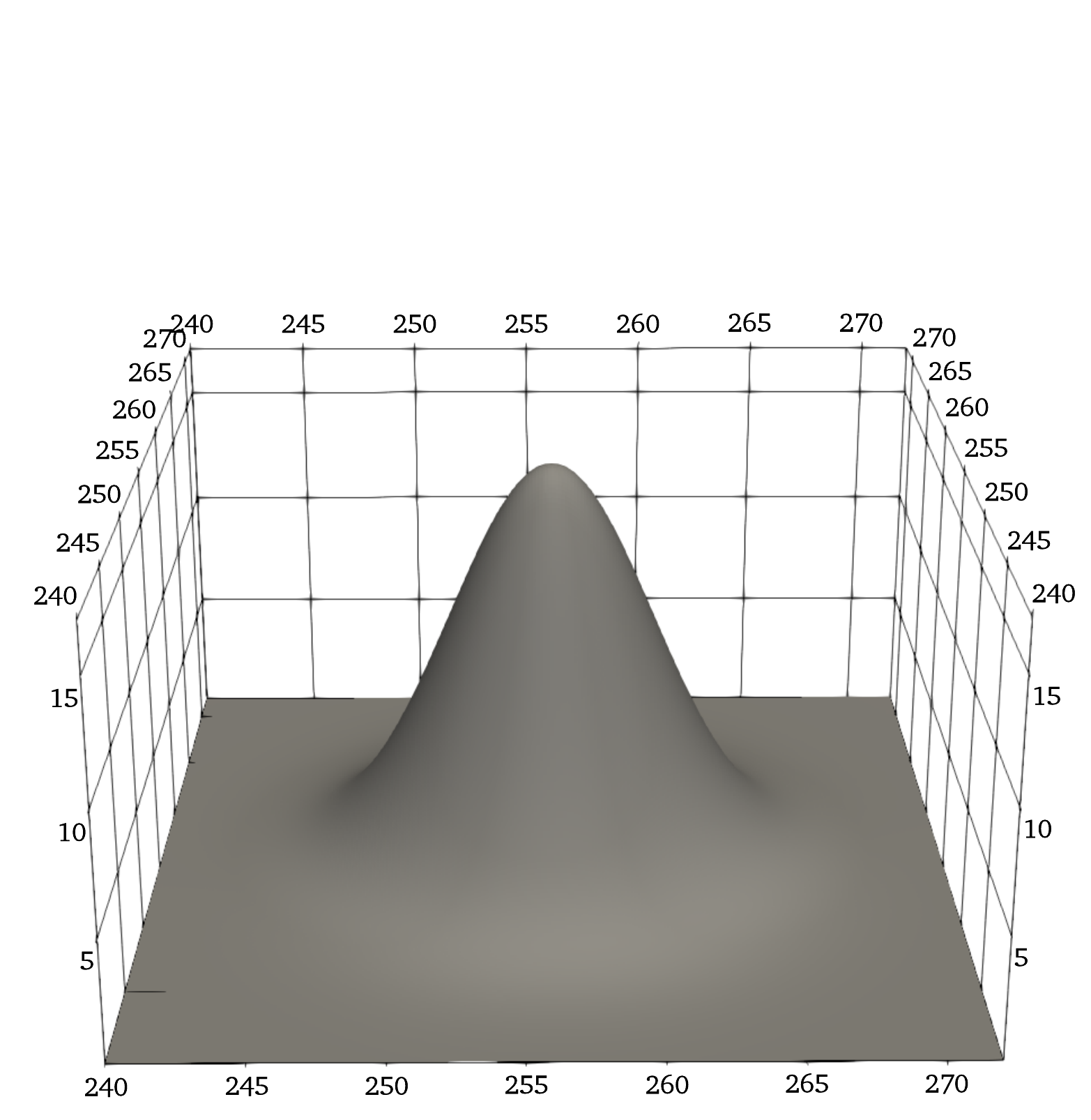}
				$\hat t=3\times10^4$
		\end{center}\end{minipage}
		
	\end{tabular}
	\caption{
		Typical examples of Q-ball decays in 2 dimensional case.
		The Q-ball is gravity-mediation type with $n=8$, $K=-0.03$ and $\xi_\tx{gravity}=0.16$.
		The horizontal axes describe spacial coordinates $\hat x$ and $\hat y$.
		The vertical axis describes an energy density.
	}
	\label{fig:ExamplesOfQballDecays2dim}
\end{figure}

We calculate the time evolution of a Q-ball for various values of $\xi$.
For small $\xi$, the Q-ball remains stable.
However, as $\xi$ increases, the charge of the Q-ball starts to decay.
We show the typical decay of a Q-ball in Fig.\ref{fig:ExamplesOfQballDecays}.
At first, the Q-charge decreases and the spacial profile starts to distort (See right figure in Fig.\ref{fig:ExamplesOfQballDecays}).
With some delay the energy starts to decrease.
After the Q-ball loses some fraction of its Q-charge, the energy becomes stable at $\hat t\sim 15\times 10^3$.
In the following calculation, we perform the numerical simulation until the energy becomes stable after they decayed and evaluate its energy and Q-charge.
We also show the results on the 2 dimensional calculation in Fig.\ref{fig:ExamplesOfQballDecays2dim}.

As a criterion of Q-ball decay, we tabulate the critical $\xi$ where its energy is reduced to 90 \% of its initial value and the Q-charge is to the 50\% when the energy density becomes stable.
The results are shown in table \ref{list:criticalzeta}.
In the gauge-mediation case, the critical value for the A-term is $\xi_\tx{gauge}^{(c)}\sim \mathcal{O} (1)$. On the other hand, in the gravity mediation case, the critical value for the A-term is $\xi_\tx{gravity}^{(c)}\sim \mathcal{O} (0.1)$.
For the gauge-mediation case, the different values of $u$ result in slightly different values of $\xi_\tx{gauge}^{(c)}$ since the mass in a vacuum is different in our calculation.

\begin{table}[htb]
	For n=4
	
	\begin{tabular}{|c|c|c|c|c| } \hline
		criteria 								   & K = $-0.03$ & K=$-0.1$		& $u=5\times 10^3$	& $u=45\times 10^3$ \\ \hline \hline
		90\% for energy 			  &      0.02	   & 	0.06			& 	1.0	  &	1.8		\\ \hline
		50\% for charge	 			   & 	0.01 		&	0.03		& 		1.5	    & 	1.4	\\ \hline
	\end{tabular}

	For n=6	
	
	\begin{tabular}{|c|c|c|c|c| } \hline
		criteria 								   & K = $-0.03$ & K=$-0.1$		& $u=5\times 10^3$	& $u=45\times 10^3$ 	 \\ \hline \hline
		90\% for energy 			  &      0.05	   & 	0.08	& 	1.5  &  	2.25\\	 \hline
		50\% for charge	 			   & 	 0.05		&	0.08		& 	1.5		&   2.25 \\ \hline
	\end{tabular}

	For n=8
	
	\begin{tabular}{|c|c|c|c|c| } \hline
		criteria 								   & K = $-0.03$ & K=$-0.1$		& $u=5\times 10^3$	& $u=45\times 10^3$	 \\ \hline \hline
		90\% for energy 			  &      0.09	   & 	0.17	&	 	1.0	  &	1.4	 \\ \hline
		50\% for charge	 			   & 	0.16 		&	0.20	& 	1.0		   &	1.4	 \\ \hline
	\end{tabular}
	
	\caption{
		The critical value of $\xi$ for the gravity-mediation type ($K = -0.03,-0.1$) and the gauge-mediation type ($u=5\times 10^3,45\times 10^3$).
	}
	\label{list:criticalzeta}
\end{table}

The upper bound of Q-charge is evaluated by $\xi^{(c)}$ as
\begin{align}
	Q_\tx{max}^\tx{(gauge)}
	&=
	\left( \frac{\sqrt{2}}{\zeta} \right)^4 
	\left( \frac{n \xi_\tx{gauge}^{(c)}}{2 |A|} \right)^{4/n}
	\left( M_*^{n-3} M_F^{4-n}m_{3/2}^{-1} \right)^{4/n}
\nonumber
\\	&=
	\left( \frac{2.5}{\zeta} \right)^4 
	\left[ 
		\frac{\xi_\tx{gauge}^{(c)}}{ |A|} 
		\left( \frac{M_*}{2.43\times10^{18}\GeV}  \right)^{n-3}
		\left( \frac{M_F}{10^6\GeV}  \right)^{4-n}
		\left(\frac{0.1\GeV}{m_{3/2}} \right)
	\right]^{4/n}
\nonumber
\\&\times
	\begin{cases}
	5\times 10^{18}
	&\quad (n=4)
	\\
	6\times 10^{28} 
	&\quad (n=6)
	\\
	6\times 10^{33}
	&\quad (n=8)
	\end{cases}
	,
	\label{eq:QmaxnonPertGauge}
\end{align}
	and 
\begin{align}
	Q_\tx{max}^\tx{(gravity)}
	&=
	 \frac{2}{|K|^{3/2}}  
	\left( \frac{n \xi_\tx{gravity}^{(c)}}{2 |A|}
	\frac{M_*^{n-3}}{m_{3/2}^{n-3}}
	 \right)^{\frac{2}{n-2}}
	 \nonumber
	\\	&=
	\left( \frac{0.03}{|K|} \right)^{3/2}
	\left[
	\frac{1}{ |A|}  \left(\frac{\xi_\tx{gravity}^{(c)}}{0.1} \right)
	\left( \frac{0.1\GeV}{m_{3/2}}\frac{M_*}{2.43\times10^{18}\GeV}  \right)^{n-3}
	\right]^{\frac{2}{n-2}}
	\nonumber
	\\&\times
	\begin{cases}
	2\times 10^{21}
	&\quad (n=4)
	\\
	3\times 10^{31}
	&\quad (n=6)
	\\
	6\times 10^{34}
	&\quad (n=8)
	\end{cases}
	\ . 
	\label{eq:QmaxnonPertGrav}
\end{align}

	\section{Evolution of Q-balls inside a neutron star}
	\label{sec_discussion}

In section \ref{sec_potential}, we classifies the three paths of the Q-ball growth, i.e. Case without $V_\tx{gravity}$, Case with $V_\tx{gravity}$ ($K<0$) and Case with $V_\tx{gravity}$ ($K>0$).
In each case, the consumption rate of neutron stars by Q-balls depends on their profile, i.e. the gauge-mediation type, the gravity-mediation type or the thin wall type.
The growth stops when the Q-ball reaches its maximum size determined by the A-term.
We consider the transition of Q-ball type during its growth, and examine whether the Q-ball consumes all baryon in the neutron star.

We also compare our conservative results with the previous perturbative discussion. (See brief summary in appendix \ref{sec_pertAterm}.)
In the following, we sometimes use the following reference values~\cite{deLavallaz:2010wp}:
\begin{align}
	\zeta&\sim 2.5 , \
	 M_F=10^3\GeV , \
	  m_{3/2}=0.1\GeV  , \ 
	  M_*=\Mpl=2.43\times 10^{18}\GeV  , \
\nonumber
\\	  \tau_\tx{NS}& =0.1\tx{Gyr} ,\
	  B_\tx{NS} =10^{57} .
\end{align}

\subsection{Case without $V_\tx{gravity}$}
\label{sec_gaugemediationGrowth}

\begin{figure}[tb]
	\begin{center}
		\includegraphics[width=1.0\textwidth ]{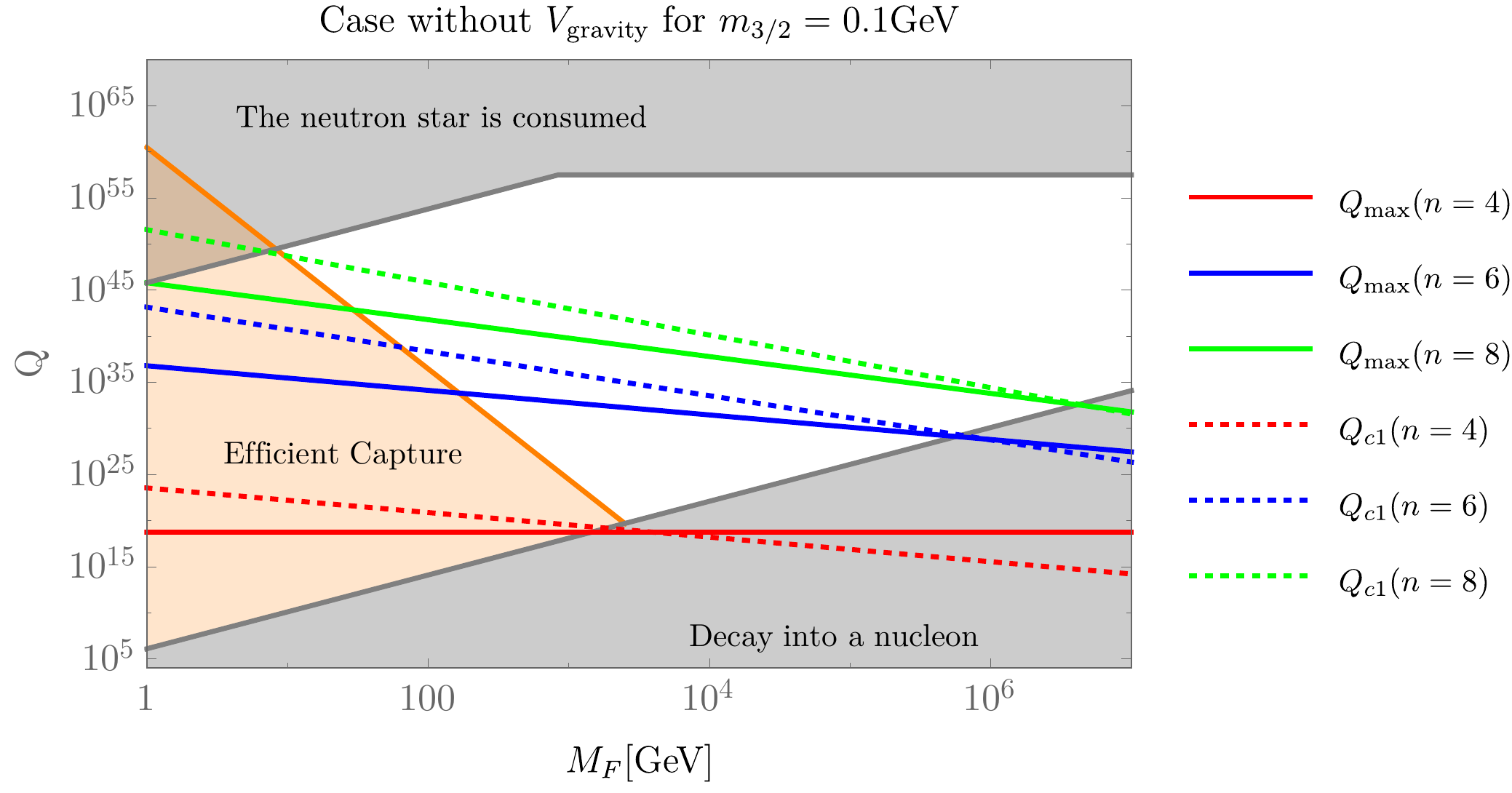}
	\end{center}
	\caption{
		 The upper limit on the Q-ball charge $Q_\tx{max}$ for the case without $V_\tx{gravity}$.
		 The orange region describes the Q-balls effectively captured by the neutron star [Eq.\eqref{eq:Qcapture}].
		 When Q-ball is captured in the neutron star, it starts to grow its Q-charge.
		 The growing Q-ball reaches its maximum size $Q_\tx{max}$ given by the A-term decay Eq.\eqref{eq:QmaxnonPertGauge} with $m_{3/2}=0.1\GeV$ (Solid color lines).
		 The color of lines means that the indices of A-term $\propto \Phi^n$ are $n=4$ (Red),  6 (Blue) and 8 (Green). 
		 In the upper gray region the Q-ball consumes all the baryons in the neutron star [Eqs. \eqref{eq:growtingGaugeQ} and \eqref{eq:QmaxGauge}].
		 In the lower gray region Q-balls decay into the nucleon [Eq.\eqref{eq:Qmin}].
		 The dotted lines describe the critical charge $Q_{c1}$ of the thin wall type Q-ball [See Eq.~\eqref{eq:gaugeTranceQc1}].
	}
	\label{fig:growthandTransition}
\end{figure}

In this case, we ignore the $V_\tx{gravity}[\Phi]$. 
Q-balls are the gauge-meditation type at first and grow to become the thin wall type, as we explained in Sec.\ref{sec_growQ}. 
We assume that the A-term suppresses the growth of Q-balls before they become thin wall type, which we confirm later. 

We show the maximum Q-ball charge $Q_\text{max}$ for $M_F= (1-10^6)\GeV$ in Fig.\ref{fig:growthandTransition}. 
DM Q-balls should satisfy the stability condition given by Eq.\eqref{eq:Qmin}, which excludes the lower gray region in the figure.
The efficient captures of Q-balls in a neutron star take place for $\gamma/\Omega\gg1$ [Eq.\eqref{eq:Qcapture}] (Orange region).
Fig \ref{fig:growthandTransition} shows that the capture process mainly occurs for the Q-balls with $M_F\lesssim 2.5\times 10^3\GeV$. 
All baryons in a neutron star are consumed by a Q-ball in the upper gray region where $B_\tx{NS}>b Q$ or 
$B_\tx{NS}>b \tau_\tx{NS} \dot Q$.
Color solid lines denote the upper limit on the Q-ball size $Q_\tx{max}$ by A-term [Eq.~\eqref{eq:QmaxnonPertGauge}]. 
Note that our analysis assumes that the Q-balls never become the thin wall type. 
Color dotted lines donate the critical charge $Q_{c1}$ of the thin wall type Q-ball [Eq.~\eqref{eq:gaugeTranceQc1}].
As you can see, $Q_\tx{max}$ is smaller than $Q_{c1}$ for stable Q-balls. 
For the broad range of $M_F$, the A-term suppresses the Q-ball growth and avoids the destruction of neutron stars. 

	\subsection{Case with $V_\tx{gravity}$ ($K<0$)}
	\label{sec_newtypeGrowth}

The $V_\tx{gravity}$ with large $m_{3/2}$ and $K<0$ results in the appearance of gravity-mediation type Q-balls.
Let us assume that the gauge-mediation type Q-balls are captured inside a neutron star.

After the capture, the Q-balls grow and start to change its profile into the gravity-mediation type when its Q-charge exceeds the critical charge~\cite{Kasuya:2000sc},
\begin{align}
	Q_\tx{eq}&\sim
	\left(  \frac{\phi_\tx{eq}}{|K|^{3/4}m_{3/2}} 	\right)^{2}
	\sim
	4\times 10^{30}
	\left( \frac{M_F}{10^6\GeV}  	\right)^{4}
	\left(   \frac{0.1\GeV}{m_{3/2}} 	\right)^{4}
	\left(\frac{0.03}{|K|}\right)^{3/2}.
\end{align}
Since the gauge-mediation type has the upper bound $Q_\tx{max}$ as described in the previous subsection, the appearance of the gravity-mediation type requires $Q_\tx{max}>Q_\tx{eq}$, which puts a condition on $m_{3/2}$ as
\begin{align}
	m_{3/2} 
	>
	8 \times 10^{-3}\GeV
	\left(\frac{ 10^{35}}{Q_\tx{max}}\right)^{1/4}
	\left( \frac{M_F}{10^6\GeV}  	\right)
	\left(\frac{0.03}{|K|}\right)^{3/8}
	\label{eq:NewtypeAppear}
\end{align}
For this parameter region the captured Q-balls grow into the gravity-mediation type. 

In this case, we need to evaluate $Q_\tx{max}$ based on the gravity-mediation type profile.
Note that the ratios of A-term and non-renormalizable term to the mass term are $\xi_\tx{gravity}\sim \phi^{n-2}/(m_{3/2}M_*^{n-3})$ and $(\phi^{n-2}/(m_{3/2}M_*^{n-3}) )^2$, both of which become order one at the same field value.
Since we found $\xi_\tx{gravity}\sim \mathcal{O}(0.1)$, the A-term suppression stops the Q-ball growth before the $V_\tx{NR}$ dominates the potential~\cite{Kasuya:2014ofa}.

We also comment on the perturbative description of $V_\tx{gravity}$.
In considering the growing Q-ball we have to take the renormalization scale independent of its charge and $\phi_0$ for a given $m_{3/2}$.
During the growth of Q-ball from $Q\sim10^{20}$ to $Q\sim 10^{40}$ in the gravity-mediation type [See Fig.\ref{fig:comparegrowt}], the field value $\phi_0$ of the Q-ball increases by about ten orders of magnitude.
This also increases the perturbative correction $K\log[\phi^2_0/(2M^2)]$ in Eq.\eqref{eq:propertygrav}.
We fix the renormalization scale M so that it does not spoil the perturbativity.
For $K=-0.03$, we can keep its contribution perturbative over the typical growth history by taking the appropriate renormalization scale
$M\sim 10^{13}\GeV\,(m_{3/2}/0.1\GeV)$.
With this choice the Q-ball has $Q= 10^{30}$ at $\phi_0 = \sqrt{2} M$ where the correction vanishes. 

Let us estimate the neutron star constraint.
The baryon consumption of gravity-mediation type given by Eq.\eqref{eq:Gravm32upperconsumption} is independent of the Q-charge. 
The constraint $B_\tx{NS} > b \tau_\tx{NS} \dot Q_\tx{grow}$ puts the lower bound on $m_{3/2}$ given by Eq.~\eqref{eq:Gravm32upperconsumption}, which is obviously satisfied in this case as seen from Eq.~\eqref{eq:NewtypeAppear}.
Thus, the new type Q-ball in this case is not constrained by the stability of neutron stars.

		\subsection{Case with $V_\tx{gravity}$ ($K>0$)}
		\label{sec_newtypeGrowthplus}

\begin{figure}[t]
	\begin{center}
		\includegraphics[width=1.0\textwidth ]{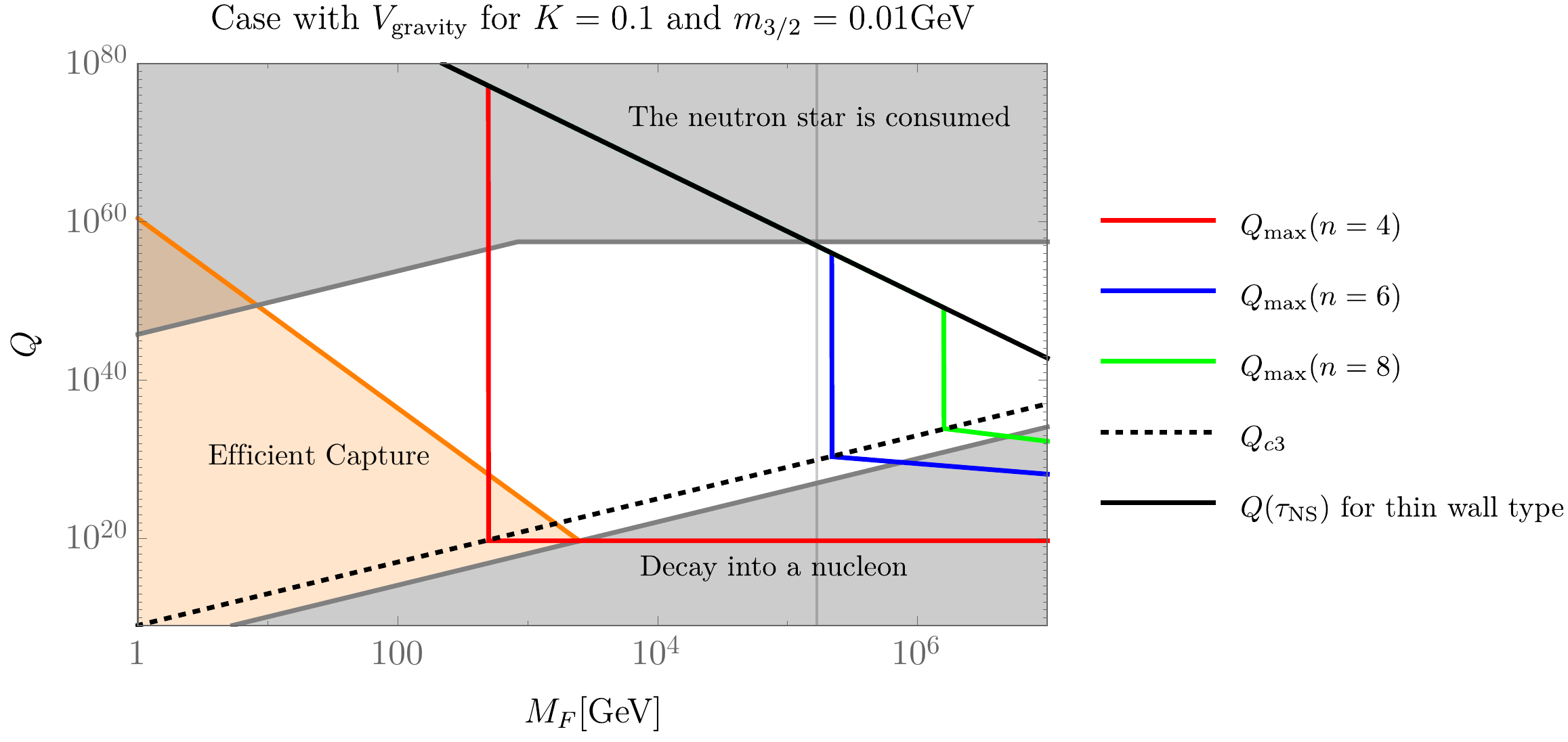}
	\end{center}
	\caption{
		The upper limit on the Q-ball charge $Q_\tx{max}$ in the case with $V_\tx{gravity}$ for $K=0.1$ and $m_{3/2}=0.01\GeV$.
		Solid color line denotes the upper bound of growing Q-balls $Q_\tx{max}$ given by Eq.\eqref{eq:QmaxnonPertGauge}.
		Above the dashed black line, the Q-balls becomes thin wall type Eq.\eqref{eq:criticalNewpos}.
		The black solid line donates the growth of the thin wall type Q-ball at $\tau_\tx{NS}$.
		The thin wall type Q-ball quickly consumes all baryon in the neutron star with $M_F<M_{F,c}$ (vertical gray line) given by Eq.\eqref{eq:criticalMfforNewpos}.
	}
	\label{fig:growthandTransitionnewpos}
\end{figure}

For $K>0$, the $V_\tx{gravity}$ does not support the Q-balls solutions.
Thus, the $V_\tx{gravity}$ makes Q-balls thin wall type when the Q-ball grows over the threshold value $Q_{c3}$ given by
\begin{align}
	Q_{c3} &
	\sim 
	\left( \frac{1 }{ \zeta} \right)^4
	\left( \frac{\phi_{c3}} {M_F} \right)^4
	\sim 
	1\times 10^{29}
	\left( 	\frac{0.1}{K} \right)^2
	\left( \frac{M_F}{10^6\GeV} \right)^4
	\left( \frac{0.1\GeV}{m_{3/2}} \right)^4,
	\label{eq:criticalNewpos}
\end{align}
where $\phi_{c3}$ is the critical value [see Eq.\eqref{eq:phic3}].
Once the Q-ball grows into the thin wall type, the field value of the Q-ball does not exceed $\phi_{c3}$ and the A-term no longer becomes effective. 
From Eq.~\eqref{eq:thinwallGrowthrate}, the thin wall type Q-ball grows over $\tau_\tx{NS}$ as 
 \begin{align}
	Q(\tau_\tx{NS})
	&\sim
	6\times 10^{76} 
	\left( \frac{M_F}{10^3\GeV}  \right)^{-8}
	\left( 	\frac{K}{0.1} \right)
	\left( \frac{m_{3/2}}{0.1\GeV} \right)^2 \ ,
\end{align}
and it consumes all the baryon in a neutron star for 
\begin{align}
	M_F < M_{F,c} 
	=3\times 10^5 \GeV 
	\left[  
	\frac{10^{57} }{B_\tx{NS} }
	\left( 	\frac{K}{0.1} \right)
	\left( \frac{m_{3/2}}{0.1\GeV} \right)^2
	\right]^{1/8}
	\ . 
	\label{eq:criticalMfforNewpos}
\end{align} 

We show the $Q_\tx{max}$ with $K>0$ and $m_{3/2} =0.01\GeV$ in Fig.\ref{fig:growthandTransitionnewpos}.
The Q-ball becomes the thin wall type above the $Q_{c3}$ (Black dashed line).
The thin wall type Q-ball quickly grows up to $Q(\tau_\text{NS})$ (Black solid line) and consumes the neutron star on $M_F<M_{F,c}$.
The color lines are the upper bound of Q-ball size $Q_\tx{max}$ [Eq.~\eqref{eq:QmaxnonPertGauge}], which does not exist for the thin wall type Q-balls. 
To avoid the constraint by the neutron star, A-term needs to suppress the Q-ball growth before it becomes the thin wall type.
Once the Q-ball is captured by the neutron star, the allowed parameter regions are $M_F\sim10^3\GeV$ for n=4 case or $M_F>M_{F,c}$ for n=6 and 8 case. 
Compared to the case without $V_\tx{gravity}$ in Fig.\ref{fig:growthandTransition}, the allowed parameter region of the case with $V_\tx{gravity}$ $(K>0)$ is small.

\subsection{Comparison to perturbative calculation}
\label{sec_comparison}

Let us compare our results with the previous calculation of the perturbative decay rate~\cite{Cotner:2016dhw}, which is summarized in appendix \ref{sec_pertAterm}.
Although they focus on the gauge mediation type, we also evaluate the perturbative decay rate for the gravity mediation type.

\begin{figure}[b!]
	\centering 
    \begin{tabular}{c}
		
		\begin{minipage}{0.5\hsize}	\begin{center}
			\includegraphics[width=0.745\textwidth]{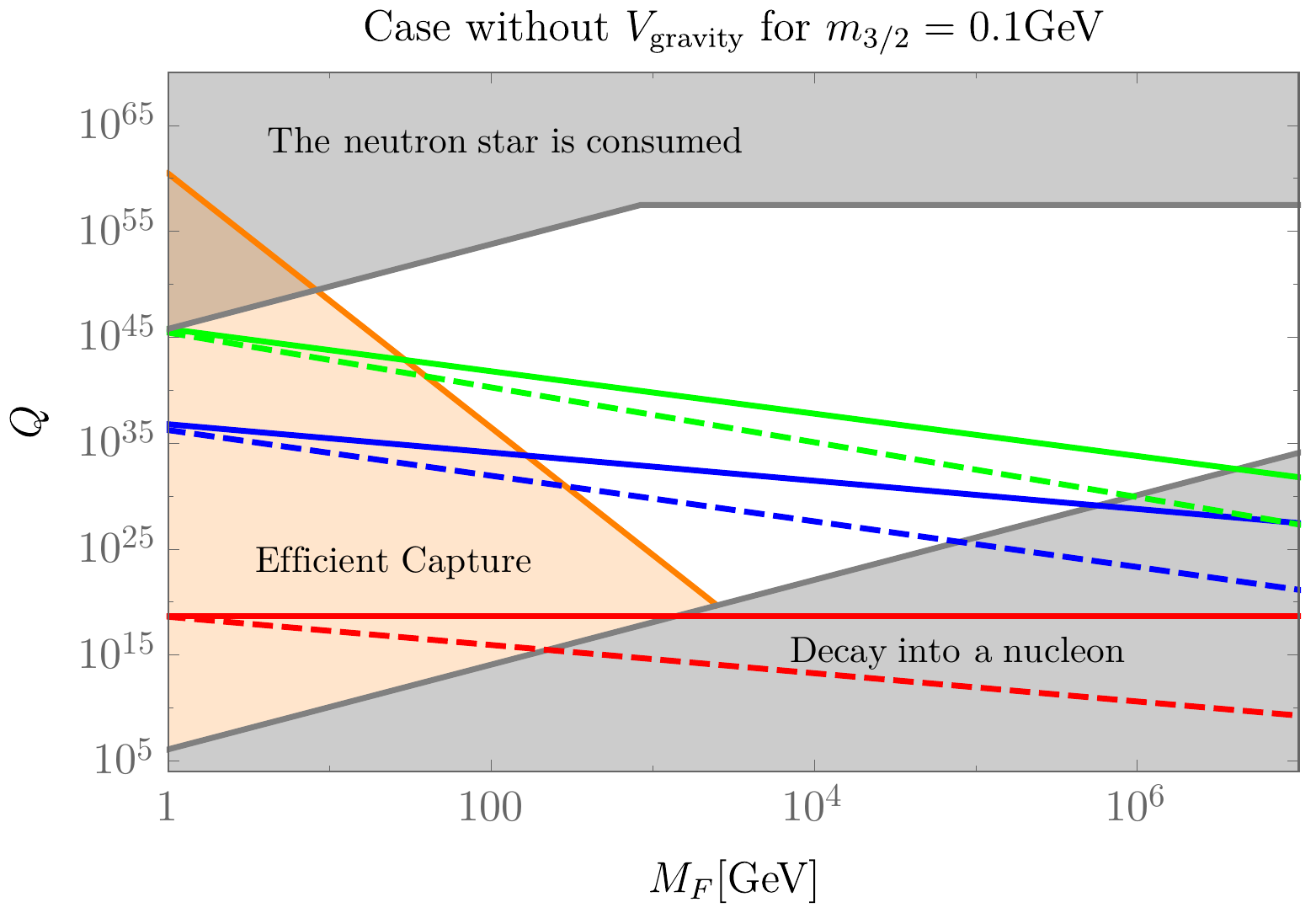}
			\end{center}\end{minipage}

		\begin{minipage}{0.5\hsize}	\begin{center}
			\includegraphics[width=1.\textwidth]{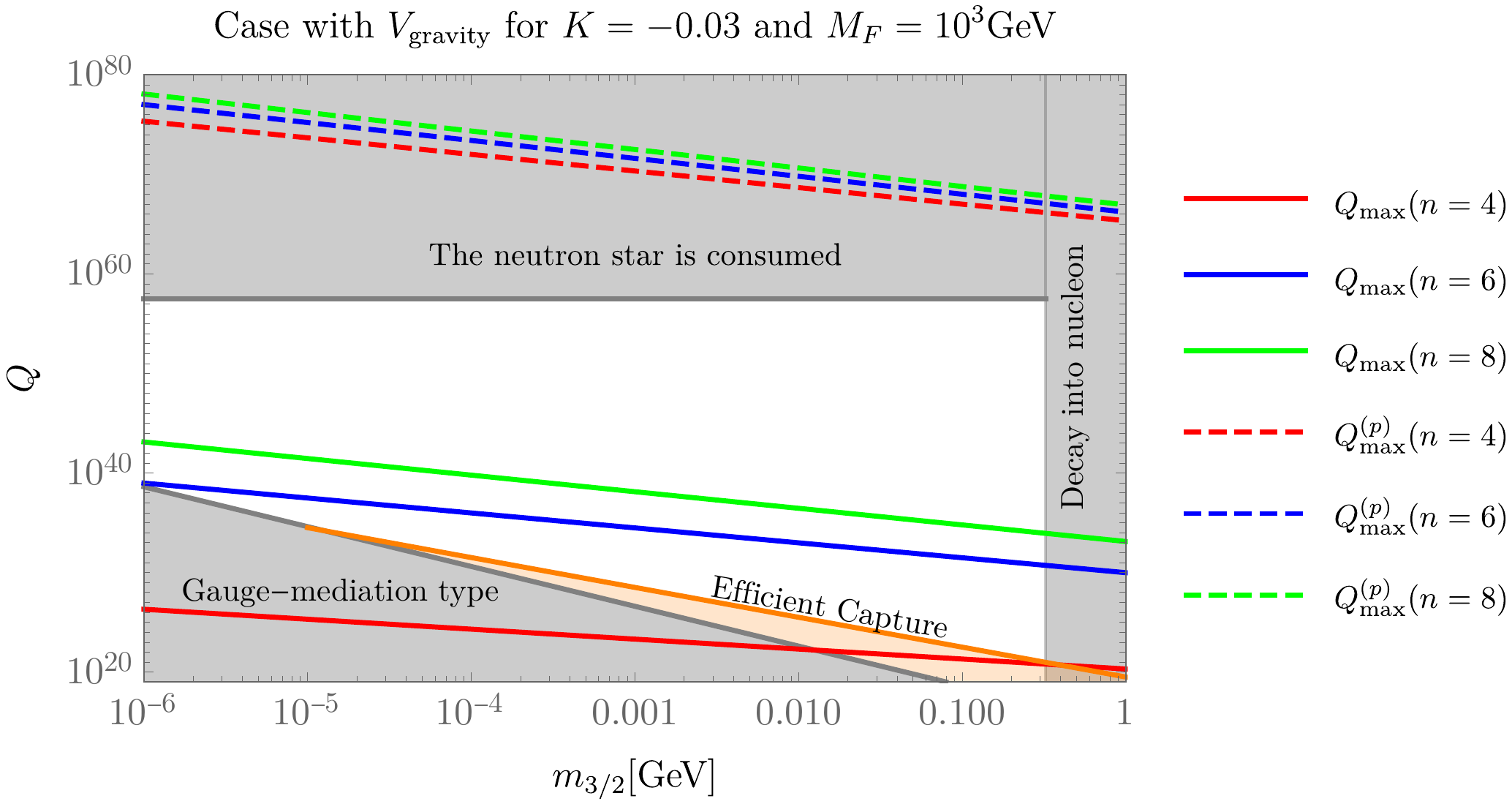}
		\end{center}  \end{minipage}
		
	\end{tabular}
	
	\caption{
	    The upper limit on the Q-ball size $Q_\tx{max}$ for our results (solid color lines) and previous calculation (dashed color lines) ~\cite{Cotner:2016dhw}.
	    The figures describe gauge-mediation type in the case without $V_\tx{gravity}$ (left) and gravity-mediation type in the case with $V_\tx{gravity}$ ($K<0$) (right).
	    The orange regions show the parameters for which Q-balls are captured by neutron stars efficiently.
		 }
	\label{fig:comparegrowt}
\end{figure}

In Fig.\ref{fig:comparegrowt}, solid color lines donate our results $Q_\tx{max}$ and dashed color lines donate the previous calculation $Q^{(p)}_\tx{max}$.

For the gauge-mediation type (left figure),
the previous calculation $Q^{(p)}_\tx{max}$ puts stronger constraints on $Q_\tx{max}$ compared to our results
[Eq.\eqref{eq:qmaxPerterbative}].
Both our and the previous results~\cite{Cotner:2016dhw} show that neutron stars are stable against the growth of Q-balls in the gauge-mediation type.

For the gravity-mediation type (right figure), interestingly, the perturbative decay [Eq.\eqref{eq:qmaxPerterbativeGrav}] is inefficient compared to our result~[Eq.~\eqref{eq:qmaxPerterbative}].
It is because the perturbative decay rate is suppressed by the momentum conservation as $\Gamma_n^{(1)}\propto \exp\left[ -(R\omega)^2n^2/(2n-2) \right]$, in which $(R\omega)^2 \simeq  4|K|^{-1} \gg 1$ in gravity mediation type.
\footnote{
	The perturbative decay may become efficient even in thin wall type when we make a detailed estimation, e.g. the exact profile of thin wall type and higher-order terms of perturbative decay. 
}

\section{Conclusion}
\label{sec_conclusion}

The Affleck-Dine mechanism can produce non-topological solitons, Q-balls, which can explain the dark matter of the universe.
The stability of neutron stars gives one of the stringent constraints on the DM Q-balls.
The DM Q-balls can be trapped inside the neutron stars, grow consuming their baryon charge, and finally destabilize the neutron stars.
In this paper, we have performed the numerical lattice simulations and confirmed that the growth of Q-balls is suppressed by the $U(1)$ breaking A-term.
Since a growing Q-ball has large field value, the A-term becomes effective and the Q-ball loses some fraction of its charge until it becomes stable against the A-term instability.

In the previous calculation, the Q-ball destabilization is evaluated by perturbative calculations for the gauge-mediation type \cite{Cotner:2016dhw} and by numerical simulations for the gravity-mediation type \cite{Kawasaki:2005xc}.
In general, growing Q-balls change the type and property.
Thus, for a comprehensive treatment of the Q-ball growth inside a neutron star, we have evaluated both the gravity and the gauge-mediation types numerically in this paper. 
We estimated the critical Q-charge above which the Q-balls start to decay.

For the gauge-mediation type (Case without $V_\tx{gravity}$), we have confirmed that the A-term effectively suppresses the Q-ball growth and DM Q-balls are not constrained.
As for the new type Q-balls (Case with $V_\tx{gravity}$), we have two possibilities for the gravity-mediation type potential, $K<0$ or $K>0$. 
In the former case, we have found that the Q-ball can become the gravity-mediation type and its growth is also suppressed by the A-term.
In the latter case, the Q-ball quickly becomes the thin wall type. 
We have found the only small parameter space, where the new type Q-balls avoids the constraint by the neutron star.

\section*{Acknowledgements}
\small\noindent
We thank Eisuke Sonomoto for allowing to use the lattice simulation code and Shinta Kasuya for a useful comment.
This work was supported by JSPS KAKENHI Grants No. 17H01131 (M. K.) and No. 17K05434 (M. K.), MEXT KAKENHI Grant No. 15H05889 (M. K.), World Premier International Research Center Initiative (WPI Initiative), MEXT, Japan (M. K., H. N.), Advanced Leading Graduate Course for Photon Science (H.N.), and the JSPS Research Fellowships for Young Scientists Grant No. 19J21974 (H. N.).

\appendix 

\section{Transition of Q-balls into the thin wall type}
\label{sec_eachhistory}

\begin{enumerate}[(I)]
	\item  Case without $V_\tx{gravity}$ :
	
	When the gravitino mass is sufficiently small, $V_\tx{gravity}$ does not dominate the potential.
	In this case, $V_\tx{gauge}$ dominates the potential for small field value $\phi<\phi_{c1}$ and $V_\tx{NR}$ does for large field value $\phi>\phi_{c1}$ with the critical value $\phi_{c1}$ given by the condition 
	\begin{align} 
    	0&=\frac{\df }{\df \phi}
    	\left( \frac{V_\tx{gauge}+V_\tx{NR}}{\phi^2}\right)
    	\bigg|_{\phi=\phi_{c1}}
    \qcn
    	\left( \frac{\Phi_{c1}}{M_*} \right)^{2n-2}
    	\left( \frac{M_*}{M_F} \right)^{4}
    	&=
    	\frac{1}{2n-4}
    	\left( 
    	[\log(\Phi_{c1}/M_\tx{mess})]^2
    	-
    	2\log(\Phi_{c1}/M_\tx{mess})
    	\right)
    	\sim O\mathcal (1) ,
    \end{align}
    which leads to
    \begin{equation}
        	\phi_{c1}\sim
    	\sqrt{2}M_* \left( \frac{M_F}{M_*} \right)^{\frac{2}{n-1}}  .
    \end{equation}
	The critical charge of Q-balls at $\phi_{c1}$ is given by
	\begin{align}
    	Q_{c1} &\sim 
    	\left( \frac{1 }{ \zeta}  \frac{\phi_{c1}}{M_F} \right)^4
    	=
    	\left( \frac{\sqrt{2} }{ \zeta} \right)^4
    	\left( \frac{M_F}{M_*} \right)^{\frac{8}{n-1}-4} 
    	.
	\label{eq:gaugeTranceQc1}
	\end{align}
	\item Case with $V_\tx{gravity}$ ($K<0$) :
	
	When gravitino mass is large enough to achieve $\phi_\tx{eq} < \phi_{c1}$, $V_\tx{gravity}$ dominates the potential for $\phi_\tx{eq}<\phi<\phi_{c2}$ and $V_\tx{NR}$ does for $\phi_{c2}<\phi$, where the critical value $\phi_{c2}$ is given by
	\begin{align} 
    	0&=
    	\frac{\df }{\df \phi}
    	\left( \frac{V_\tx{gravity}+V_\tx{NR}}{\phi^2}\right)
    	\bigg|_{\phi=\phi_{c2}}
    	~~\Rightarrow ~~ 
    	\phi_{c2}
    	=
    	\sqrt{2} M_* \left( \sqrt{\frac{|K|}{n-2}}\frac{m_{3/2}}{M_*} \right)^{\frac{1}{n-2}} 
    	\ .
	\end{align}
	The critical charges of Q-balls for $\phi_{c2}$ is
	\begin{align}
    	Q_{c2} &\sim 
    	\left( \frac{\phi_{c2}} {|K|^{3/4}m_{3/2}} \right)^2
    	\sim 
    	\left( \frac{ \sqrt{2} M_*}{ |K|^{3/4}m_{3/2} } \right)^2
    	\left( \frac{|K|^{1/2}m_{3/2}}{\sqrt{n-2}M_*} \right)^{\frac{2}{n-2}} 
    	.
	\end{align}
	\item Case with $V_\tx{gravity}$ ($K>0$) :
	
	With $K>0$, $V_\tx{gravity}$ does not support the condition of the Q-ball. 
	Then, the thick wall solution appears for $\phi\sim \phi_{c3}$ given by
	\begin{align} 
    	0&=
    	\frac{\df }{\df \phi}
    	\left( \frac{V_\tx{gravity}+V_\tx{gauge}}{\phi^2}\right)
    	\bigg|_{\phi=\phi_{c3}}
    	\qcn
    	\phi_{c3}
    	&=
    	\sqrt{2}\sqrt{
    		\frac{M_F^4}{K m_{3/2}^2}
    		\left( 
    		[\log(\Phi_{c3}/M_\tx{mess})]^2
    		-2\log(\Phi_{c3}/M_\tx{mess})
    		\right)
    	}
    	\sim 
    	\sqrt{2}\frac{M_F^2}{K^{1/2} m_{3/2}}
    	\label{eq:phic3}
    	,
    	\\
    	Q_{c3} &\sim 
    	\left( \frac{1}{ \zeta} \right)^4
    	\left( \frac{\phi_{c3}} {M_F} \right)^4
    	\sim 
    	1\times 10^{29}
    	\left( 	\frac{0.1}{K} \right)^2
    	\left( \frac{M_F}{10^6\GeV} \right)^4
    	\left( \frac{0.1\GeV}{m_{3/2}} \right)^4
    	\label{eq:criticalQfornewKplus}
	\end{align}
	In this case, Q-balls smoothly change from the gauge-mediation type to the thin wall type.
\end{enumerate}

\section{Perturbative decay of Q-ball by A-term}
\label{sec_pertAterm}

Let us summarize the perturbative decay rate of Q-ball by A-term \cite{Cotner:2016dhw}.

The decay rate into N particles  is given by \cite{Cotner:2016dhw}
\begin{align}
\Gamma_n^{(N)}
&=4\pi^5 2^{n-2N}  |g_n|^2 n^N R^6 \Phi_0^{2(n-N)} \omega^{2N-1 } J^{(N)}_n
\end{align}
with $g_n= \frac{m_{3/2}}{M_*}\frac{A }{nM_*^{n-4}}$ in this case.
The decay rate depends on N as
$\Gamma_n^{(N)}\propto \left( \frac{\omega}{2\Phi_0} \right)^{2N}  \propto 	Q^{-N}$.
Thus, with large Q, we only need to focus on $N=1$ scattering. 
The phase space integral for $N=1$ is a function of $R\omega$:
\begin{align}
J^{(1)}_n(R\omega)
&=
\exp\left( -\frac{R^2\omega^2  }{2(n-1)} (n^2 -(4\pi R\omega)^{-2}) \right)
\frac{ 	\sqrt{ (n^2 -(4\pi R\omega)^{-2}) } }{(2\pi)^2}
\theta(n -(4\pi R\omega)^{-1}) 
.
\end{align}
In this paper, we use the following decay rate as the perturbative estimation:
\begin{align}
\Gamma_n^{(1)}
&=2\pi^5J^{(1)}_n(R\omega)
\omega^1
\left( \frac{m_{3/2}}{M_*} \right)^2
\left( \frac{A }{\sqrt{n}} \right)^2 
(RM_*)^6 
\left(\frac{\sqrt{2}\Phi_0}{M_*}\right)^{2(n-1)} .
\end{align}
$\Gamma_n^{(1)}$ represents the decay rate of Q-charges into the phonon on the Q-ball with energy $\Delta E_{ex} = n \omega$. 
The phonon may decay into some light particles, but it is unclear what the decay channel is and whether the inverse process is negligible.
In this paper, we do not study the detail of this process.

For gauge-mediation type, the maximum size of Q-balls is given by the balance of the Q-ball growth $\dot Q_{grow}\propto Q^{1/2}$ and the decay rate $\Gamma_n^{(1)}\propto Q^{(2n+3)/4} $ :
\begin{align}
    Q_\tx{max}^\tx{(gauge:p)}
    &=
    \left( 
        \left( \frac{10^6\GeV}{M_F} \frac{2.5}{\zeta}  \right)^{2n-5}
        \left( \frac{\Mpl}{2.43\times10^{18}\GeV}  \right)^{2n-6}
        \left(\frac{0.1\GeV}{m_{3/2}} \right)^2
        \left(\frac{1}{|A|} \right)^2
    \right)^{\frac{4}{1+2n}}
    \\&\quad\times
    \begin{cases}
    4\times 10^{10}
    &\quad (n=4)	\\
    2\times 10^{23} 
    &\quad (n=6)	\\
    8\times 10^{29} 
    &\quad (n=8)
    \end{cases}
    \label{eq:qmaxPerterbative}
\end{align}

For the gravity-mediation type, the maximum size is given by
\begin{align}
	&Q_\tx{max}^\tx{(gravity:p)}
	=
	\left(  
	\frac{ 10^{-4} \GeV 4|K|^{-1} \left(  \frac{1\GeV}{m_{3/2}}\right)^2 }{\Gamma_n^{(1)}(Q=1)}
	\right)^{\frac{1}{n-1}}
	\\	&=
	\left( 
	\left( \frac{\Mpl}{2.43*10^{18}\GeV}  \right)^{2n-6}
	\left(\frac{0.1\GeV}{m_{3/2}} \right)^{2n-3}
	\frac{1}{|A|^2}
	\right)^{\frac{1}{n-1}}
	\times
	\begin{cases}
	3\times 10^{30}
	&\quad (n=4,K=-0.1)	\\
	2\times 10^{38} 
	&\quad (n=6,K=-0.1)	\\
	4\times 10^{41} 
	&\quad (n=8,K=-0.1)\\
	1\times 10^{67}
	&\quad (n=4,K=-0.03)	\\
	1\times 10^{68} 
	&\quad (n=6,K=-0.03)	\\
	6\times 10^{68} 
	&\quad (n=8,K=-0.03)
	\end{cases}
	.
	\label{eq:qmaxPerterbativeGrav}
\end{align}
Since the momentum conservation suppresses the decay into phonons, the A-term is less effective for the Q-balls in the gravity-mediation type .

\small
\bibliographystyle{utphys}
\bibliography{RefQballdec}
\end{document}